\documentclass[11pt,a4paper]{article}
\pdfoutput=1
\usepackage{jheppub}
\usepackage{mathrsfs,graphicx,rotating,amsmath,amsfonts,mathtools,booktabs,wasysym,caption}
\usepackage{hyperref}
\usepackage{slashed}
\usepackage[table,xcdraw,dvipsnames]{xcolor}
\usepackage{graphicx}
\usepackage{bbold}
\usepackage[utf8x]{inputenc}
\usepackage[english]{babel}
\usepackage{multirow,multicol}
\usepackage{epstopdf}
\usepackage{bbm}
\usepackage{changepage}
\usepackage{appendix}
\usepackage{systeme}
\usepackage{libertine}
\usepackage{braket}
\usepackage{wasysym}
\usepackage{empheq}
\usepackage{cancel}
\usepackage{enumitem}
\usepackage{mathrsfs}
\usepackage{lmodern}
\usepackage{tabularx}
\usepackage{multicol}
\usepackage{color}
\usepackage{mathtools}
\usepackage{verbatim}
\usepackage{amssymb}
\usepackage{amsfonts}
\usepackage{arydshln}

\newcommand{\be}{\begin{equation}}
\newcommand{\ee}{\end{equation}}
\newcommand{\bea}{\begin{eqnarray}}
\newcommand{\eea}{\end{eqnarray}}

\newcommand{\mc}{\mathcal}

\newcommand{\beqa}{\begin{eqnarray}}
\newcommand{\eeqa}{\end{eqnarray}}

\setcitestyle{square}

\DeclarePairedDelimiter\abs{\lvert}{\rvert}%
\title{Catch-Me-If-You-Can: The Overshoot Problem and the Weak/Inflation Hierarchy}
\author[a]{Joseph P. Conlon,} \author[a]{Filippo Revello}
\affiliation[a]{Rudolf Peierls Centre for Theoretical Physics\\ Beecroft Building, Clarendon Laboratory, Parks Road, University of Oxford, OX1 3PU, UK}
\emailAdd{joseph.conlon@physics.ox.ac.uk}\emailAdd{filippo.revello@physics.ox.ac.uk}
\abstract{We study the overshoot problem in the context of post-inflationary string cosmology (in particular LVS). LVS cosmology features a long kination epoch as the volume modulus rolls down the exponential slope towards the final minimum, with an energy density that scales as $m_s^4$.  It is a known fact that such a roll admits attractor tracker solutions, and if these are located the overshoot problem is solved. We show that, provided a sufficiently large hierarchy exists between the inflationary scale and the weak scale, this will always occur in LVS as initial seed radiation grows into the tracker solution. The consistency requirement of ending in a stable vacuum containing the weak hierarchy therefore gives a preference for high inflationary scales -- an anthropic argument, if one likes, for a large inflation/weak hierarchy.  We discuss various origins, both universal and model-dependent, of the initial seed radiation (or matter). One particularly interesting case is that of a fundamental string network arising from brane inflation -- this may lead to an early epoch in which the universe energy density principally consists of gravitational waves, while an LVS fundamental string network survives into the present universe.}
\begin{document}

\maketitle

\section{Introduction and the Overshoot Problem}
\label{intro}

Inflation is (justly) widely believed to be the physics underlying the origin of structure in the universe, through quantum fluctuations in a primordial inflaton field during a quasi-de Sitter phase in the early universe. While the precise energy scales involved in inflation are not known, in most models these are high. $V_{inf} \sim \left( 10^{15} {\rm GeV} \right)^4$ is typical, 
while even larger values arise in high-scale inflation models associated to observable levels of primordial gravitational waves. For a review of the physics of inflation, see \cite{BaumannBook}. We shall assume without further comment that the universe did indeed go through an inflationary phase in its earliest moment.

Whatever the universe was, today its vacuum state involves an electroweak symmetry broken at a characteristic energy scale $E_{weak} \sim 100 \, {\rm GeV}$. Despite the discovery of the Higgs boson in 2012, the origin of this scale is not understood. It appears unstable to quantum corrections, requiring either fine tuning or the appearance of additional sectors or symmetries (for example, supersymmetry) to control radiative corrections and maintain a Higgs mass and vev sixteen orders of magnitude below the Planck scale.

String theory is our leading candidate for a fundamental theory of nature that includes quantum treatments of both gauge and gravitational interactions. If string theory is the theory of \emph{this} world, it must account for a cosmological evolution that started with an inflationary phase and ends in a vacuum state today characterised by $E_{weak} \ll E_{inflation}$ which also includes the many other hierarchies of particle physics (for example, the Yukawa couplings).

To even talk about string vacua, we are required to confront the physics of moduli stabilisation and construct models which both stabilise moduli and exhibit a stable (ideally de Sitter) minimum. Although many aspects of our discussion extend more generally, here we will mostly consider type IIB compactifications for which the most developed models are KKLT and LVS \cite{kklt, bbcq, cqs}.  However, all string theories have a vanishing energy in the decompactification limit ($\mc{V} \to \infty$ or $g_s \to 0$). String models therefore have a natural tendency to exhibit runaway behaviour from the interior of moduli space towards the boundary, and so all vacua with (effectively) zero vacuum energy will contain a barrier height to decompactification. 

Although precise details are model-dependent, the barrier height is normally much smaller than the inflationary energy scales (as the barrier is associated to the vacuum and \emph{not} the physics during inflation). In a typical model with $m_{3/2} \sim 100\, {\rm TeV}$, the barrier height may be $V_{barrier} \sim m_{3/2}^2 M_P^2 \sim 10^{-26} M_P^4$ \cite{Kallosh:2004yh,Blanco-Pillado:2005arr,Kallosh:2007wm} (or $V_{barrier} \sim m_{3/2}^3 M_P$ in the case of LVS \cite{Conlon:2008cj}). For an inflationary scale $V_{inf} \sim \left( 10^{15} {\rm GeV} \right)^4$, this gives
\be
\frac{V_{barrier}}{V_{inf}} \sim 10^{-13}.
\ee 
This leads to the \emph{overshoot} problem \cite{BrusteinSteinhardt} - how can the minimum `trap' the post-inflationary theory in the vacuum? To make the problem more graphic: suppose a ball rolls from the top of Mount Everest towards a hole with vertical sides of one nanometer. How do we trap the ball in the hole?

Neglecting other sources of energy density, the equations of motion for a scalar field $\phi$ in an expanding universe are
\begin{eqnarray}
\ddot{\phi} + 3H\dot{\phi} & = & - \frac{\partial V}{\partial \phi}, \\
H^2 & = & \frac{1}{3} \left( V(\phi) + \frac{\dot{\phi}^2}{2} \right).
\end{eqnarray}
After inflation occurs at a scale $\Lambda_{\rm{inf}}$, the field starts rolling down its potential. String theory potentials are naturally steep. Potentials that are power-law in a volume are exponential in the 
canonical field (e.g. LVS, with $V(\Phi) \sim \exp \left( - \sqrt{\frac{27}{2}} \Phi \right)$) while non-perturbative potentials (as in KKLT) are double exponentials, $V(\Phi) \sim \exp \left( - \alpha \exp \left( \beta \Phi \right) \right)$. The initial potential energy is converted into kinetic energy of the scalar field and the universe enters a kination phase, where $\dot{\Phi}^2 \gg V(\Phi)$ and energy density scales as $\rho_{\Phi} \sim a^{-6}$. If no other sources of energy density are present,  the field will approach the minimum of the potential still in the kination phase. In this case, the field kinetic energy will be much greater than the barrier height and the field will overshoot the barrier and run away to the decompactification limit at infinity. This is the \emph{overshoot} problem.

Although it appears obvious, we note explicitly here that in this conventional formulation of the problem, 
the overshoot problem is made worse either by (a) making the inflationary scale larger or (b) reducing the barrier height. We emphasise this explicitly because one of the key points of this paper is that this conventional wisdom should be inverted in LVS.

 The overshoot problem has been periodically studied in the literature on string cosmology, and solutions to it can be placed in two classes. In the first, arguably less interesting, class the problem is solved by removing any hierarchy between the inflationary scale and the barrier height. This can be done either by keeping the barrier height low and associated to the physics of the weak scale but working with models of very low-scale inflation (for example, see \cite{German:2001tz}), or alternatively by continuing with high-scale inflation and bringing the barrier up to a similar scale (for example, see \cite{Kallosh:2004yh}). This latter approach is associated with models that regard the weak scale as an accidental feature generated by fine-tuning, when there is no \emph{dynamical} reason for the hierarchical smallness of the weak scale.

The second, arguably far more interesting, class of solutions uses dynamical effects to avoid overshooting ( used, for example, in \cite{Barreiro:1998aj, hepth0001112, hepph0010102, hepth0408160, hepph0506045, Conlon:2008cj}). These exploit the fact that, 
as $\rho_{kination} \sim a^{-6}$, almost any other early source of energy $\rho_{extra}$ will eventually catch up with the kinetic energy of the scalar (as e.g. $\rho_m \sim a^{-3}$ and $\rho_{\gamma} \sim a^{-4}$).
\begin{equation}
H^2 = \frac{1}{3 M_P^2}\Big(\rho_{extra}+ \frac{1}{2}\dot{\Phi}^2+ V(\Phi) \Big) ,
\end{equation}
The  $\rho_{extra}$ term increases the Hubble friction to a level where it can halt the runaway behaviour.
Provided $\rho_{extra}$ catches up with the kinetic energy before the field reaches the barrier, 
the modulus will settle into the minimum and avoid both overshooting and decompactification. 

In the next section (Section \ref{sec:kination}) we review the dynamics of a kination epoch (while the field is fast-rolling down the exponential slope) in a language adapted to string moduli with runaway exponential potentials. Whereas previous studies have already exploited the presence of initial radiation to solve the overshoot problem, less attention has been devoted to the possible origins of the latter. In Section \ref{sc:rad}, we discuss some model-independent ways to generate initial radiation in a typical compactification scenario (with LVS as a benchmark example), along with some phenomenological implications. These are further developed in Section \ref{sec:ph}, where we argue that that if these generic sources of radiation are to solve the overshoot problem in LVS, a large hierarchy must exist between the inflationary and electroweak\footnote{Taken to be close to the scale of susy breaking.} scales.

\section{Kination}
\label{sec:kination}

Soon after inflation ends, the field starts rolling down from the high inflationary scale and enters the (LVS) exponential slope. It rapidly converts potential into kinetic energy and the universe enters a period of kination domination. In this epoch,
the energy density of the Universe is dominated by the kinetic energy of the rolling scalar and so the potential energy is, to first approximation, irrelevant.

\subsection{Pure Kination Epoch}

In a pure kination epoch, where we neglect all other sources of energy, the scalar field equations of motion
\begin{eqnarray}
\ddot{\phi} + 3H\dot{\phi} & = & - \frac{\partial V}{\partial \phi}, \\
H^2 & = & \frac{1}{3 M_P^2} \left( V(\phi) + \frac{\dot{\phi}^2}{2} \right),
\end{eqnarray}
reduce to 
\begin{eqnarray}\label{HubbleEq1}
\ddot{\phi} + 3H\dot{\phi} & = & 0, \\
H^2 & = & \frac{\dot{\phi}^2}{6 M_P^2}.
\label{HubbleEq2}
\end{eqnarray}
The resulting equation for $\phi$,
\be
M_P \, \ddot{\phi} + \sqrt{\frac{3}{2}}  \dot{\phi}^2 = 0,
\ee
is solved by (denoting the initial condition $\phi(t_0)) = \phi_0$. 
\be
\label{fieldkination}
\phi = \phi_0 + \sqrt{\frac{2}{3}} M_P \ln \left( \frac{t}{t_0} \right).
\ee
One other integration constant has been removed by requiring that, as conventional, a time coordinate of $t=0$ corresponds to a (formal) initial singularity where energy densities diverge. The initial time $t_0$, assumed to be the beginning of the kination epoch, is determined by requiring the kinetic energy to dominate over the potential one:
\begin{equation}\label{eq:to}
\frac{M_P^2}{3 t_0^2} \gtrsim V (\phi_0)\sim \Lambda^4_{\rm{inf}}.
\end{equation} 
For $t < t_0$, the approximations \eqref{HubbleEq1}-\eqref{HubbleEq2} are no longer justified. The scale factor behaves as
\be
a(t) \propto t^{1/3},
\ee
which follows immediately from $H^2 \equiv \frac{\dot{a}(t)^2}{a(t)^2} = \frac{\dot{\phi}^2}{6 M_P^2}$. The energy density during a kination epoch therefore drops off as
\be
\rho_{kination}(t) \propto \frac{1}{a(t)^6}.
\ee

Note that in a kination epoch, the field evolves through approximately one Planckian distance each Hubble time. From a stringy perspective, this is automatically interesting: transPlanckian field excursions require a theory of quantum gravity in order to ensure control of the effective field theory over such large displacements. While this result is long-standing, it has received more recent attention following the Swampland Distance Conjecture \cite{hep-th/0605264, 181005506}. It automatically follows that during an
extended kination epoch lasting many Hubble times, the field must traverse a substantially transPlanckian distance.
Assuming the initial inflationary epoch occurred towards the centre of moduli space, any long kination epoch propels the system towards the boundary.

In general, there are control issues for field excursions $\Delta \Phi \gg M_P$. Fortunately, in the LVS context we are primarily considering, the interpretation of the field excursion is straightforward - the field $\Phi$ controls the overall volume, and the transPlanckian excursion corresponds to a growth in the size of the extra dimensions as the field rolls towards the decompactification limit. While, as per the distance conjecture, there is a tower of states that becomes `light' (here, the KK modes) this does not affect control -- which actually improves as the volume increases, as the scale of the potential drops even faster and we move deeper into the supergravity limit. While the above is true in a static limit, slightly more care should be taken in a time dependent background, where energy is not conserved and the cutoff can vary with time. In general, an EFT with a time dependent cutoff $\Lambda(t)$ is still well defined if the mixing between the high and low (with respect to the cutoff) energy modes can be neglected. Quantitatively, this amounts to the requirement that variations in the cutoff scale be adiabatic \cite{Baumann:2014nda}, \emph{i.e}
\begin{equation}\label{eq:lim}
\left| \frac{d \Lambda(t)}{d t } \right| \ll \Lambda(t)^2.
\end{equation}
If we take the cutoff to be the KK scale, 
\begin{equation}
\Lambda (t) = \frac{M_P}{\mathcal{V}_0^{2/3}}  \left( \frac{t_0}{t} \right)^{2/3}, \quad \quad  \frac{\abs{\dot{\Lambda}}}{\Lambda^2}  = \frac{2}{3 M_P}\left( \frac{\mathcal{V}_0^2}{t_0^2\, t} \right)^{1/3} \ll 1 \quad \text{for} \quad t > t_0,
\end{equation}
so that in our example \eqref{eq:lim} is satisfied at any time during the evolution.\footnote{Already for $t=t_0$,  $\abs{\dot{\Lambda} / \Lambda^2} \sim (\Lambda_{\rm{inf}}/M_P)^{10/9} \ll 1$.}
Moreover, the dependence of the various mass scales on the volume is such that there are no instances of \emph{level-crossing}, where states either appear or disappear from the EFT as the cutoff varies.

Let us relate this kination epoch to the compactification parameters. The basic kinetic term in IIB compactifications originates from
\be
K = - 3 \ln \left( T + \bar{T} \right),
\ee
where $T = \tau + i c$ is a K\"ahler modulus whose real part corresponds to the volume of a 4-cycle, $\tau \sim \mc{V}^{2/3}$, where $\mc{V}$ is the overall Calabi-Yau volume, and all volumes are measured in appropriate powers of $l_s = 2 \pi \sqrt{\alpha'}$. The imaginary part $c$ is an axion from an RR 4-form. The canonical field $\Phi$ is
\be
\Phi = M_P \sqrt{\frac{3}{2}} \ln \tau \equiv M_P \sqrt{ \frac{2}{3}} \ln \mc{V},
\ee
It follows that for a field excursion $\Delta \Phi = \langle \Phi \rangle - \langle \Phi_0 \rangle $, we can write
\be
\sqrt{\frac{2}{3}} M_P \ln \left( \frac{t}{t_0} \right) = \Delta \Phi = \langle \Phi \rangle - \langle \Phi_0 \rangle 
= \sqrt{ \frac{2}{3}} M_P \ln \left( \frac{\mc{V}}{\mc{V}_0} \right).
\ee
In this stringy kination epoch, it follows that the evolution of the internal compactification volume is
linear in cosmic time,
\be
\mc{V} = \mc{V}_0 \left( \frac{t}{t_0} \right).
\ee
In the context of LVS, as the terms contributing to the potential energy in LVS all behave as
$V \sim \mc{V}^{-3}$, this shows that during the kination-dominated roll down the exponential slope, the magnitude of the LVS potential energy behaves as
\be
V_{LVS, kination}(t) \sim t^{-3}.
\ee
We note that (as must be the case) this also follows from using the exponential potential
$V \propto \exp \left( - \sqrt{\frac{27}{2}} \Phi \right)$ and the time-evolution of the canonical field $\Phi$,
\be
V_{LVS}( \Phi) \sim \exp \left( - 3 \ln t \right) \sim t^{-3}.
\ee
As the overall energy density $V_{total} \sim t^{-2}$, the potential energy in this epoch is therefore suppressed compared to the overall energy by a factor $\frac{t_0}{t}$. This self-consistently justifies our decision to drop the potential in the analysis of the kination epoch. As a last important point, let us notice that since the string scale is $m_s \sim M_P/\sqrt{\mathcal{V}}$,
\begin{equation}
\rho_{\rm{kin}} = \frac{M_P^2}{3 t^2} \sim m_s^4
\end{equation}
for the whole of the kination epoch. 

The appearance of such a high energy density appears striking as it appears to raise potential questions about the validity of the EFT (given the overall energy density is higher than $m_{KK}^4$, and KK modes are excluded from the 4d effective field theory). However, we note that the relevant quantity controlling the validity of EFTs in a cosmological setting is normally the Hubble scale ($H \sim \left( \rho/M_P^2 \right)^{1/2}$ and not simply $\rho^{1/4}$), as this relates to the scales of actual dynamical processes within the cosmological background (we are not studying, for example, particle scattering at energies $E > M_{KK}$). Indeed, energy densities comparable to (or larger than) $\rho \sim m_s^4$ are common in various well-studied scenarios of stringy inflation, for example axion monodromy inflation \cite{Silverstein:2008sg,McAllister:2008hb} or brane inflation \cite{Dvali:1998pa,Burgess:2001fx,Dvali:2001fw,Sarangi:2002yt, Kachru:2003sx,Baumann:2007ah}. In our example, the Hubble and cutoff scales (which we take to be the KK scale) are proportional to 
\begin{equation}
H(t) \sim \frac{m_s^2}{M_P} \sim \frac{1}{t}, \quad \quad \text{and} \quad \quad m_{KK}(t)  \sim \frac{1}{t^{2/3}},
\end{equation}
and so $m_{KK}(t) \gg H(t)$, suggesting that the EFT should be under good control. 

Nonetheless, this argument notwithstanding, we do note that with any energy density greater than the KK scale, one can reasonably be concerned that there could exist some instability or way in which the KK (or string) modes manifest themselves and either modify or lead to novel phenomena inside the 4d low-energy effective field theory. Any such effects would have the potential to invalidate the standard understanding of reheating in string cosmology (both in this scenario, and also in other scenarios such as brane inflation or axion monodromy).

As well as these possible downsides, given the universal coupling of the volume, it is also tempting to speculate whether the presence of such a high energy density might be able to excite a (very small number) of KK or stringy modes, and whether this could carry any observable consequences. A possible mechanism could involve cosmological particle production in a time-varying background, but we leave such an investigation to future work.

\subsection{The Tracker Solution}

We here review the well-known existence of an attractor tracker solution for exponential potentials \cite{Wetterich:1987fm,Copeland:1997et,Ferreira:1997hj}, which implies that, in the presence of sufficient initial radiation (or matter), the overshoot problem can be solved \cite{Barreiro:1998aj, hepth0001112, hepph0010102, hepth0408160, hepph0506045, Conlon:2008cj}. 

The tracker solution relies on additional contributions that redshift slower than kinetic energy.
For a generic cosmic fluid with equation of state $P= (\gamma - 1) \rho$,  $\rho \sim a ^{-3 \gamma}$, and so this condition is equivalent to $\gamma < 2$. Both matter and radiation satisfy this condition. Given the high inflationary scales, the presence of stable matter at the end of inflation appears unlikely (although we consider primordial black holes later).  We mostly consider just the case of initial radiation, where $\rho_{extra} = \rho_{\gamma}$ (note we use $\rho_{\gamma}$ to denote any form of radiation, not just photons).

For this system, the Friedmann equations are
\bea
\dot{H} & = & - \frac{1}{2 M_P^2}\Big(\rho_{\gamma}+P_{\gamma} + \dot{\Phi}^2\Big) = - \frac{1}{2 M_P^2} \Big( \gamma \rho_{\gamma}+ \dot{\Phi}^2\Big), \\
H^2 & = & \frac{1}{3 M_P^2}\Big(\rho_{\gamma}+ \frac{1}{2}\dot{\Phi}^2+ V(\Phi) \Big) ,
\eea
while the energy conservation equation is
\begin{equation}
\dot{\rho}_{\gamma}= -3 H \big( \rho_{\gamma}+P_{\gamma} \big) = -3 H \gamma \rho_{\gamma}.
\end{equation}
Alternatively, one can switch to the variables (see \cite{Copeland:1997et})
\begin{equation}
x = \frac{\dot{\Phi}}{M_P } \frac{1}{\sqrt{6} H}, \quad \quad y = \sqrt{\frac{V(\Phi)}{3}} \frac{1}{M_P H},
\end{equation}
which encode the fractional energy densities in kinetic and potential energy respectively, $\Omega_k = x^2,\Omega_p = y^2$,  with a radiation energy density $\Omega_{\gamma}= 1- x^2 -y^2$.
The same dynamical evolution can be recast as the system
\begin{equation}\label{eq:xy2}
\left\{ \begin{aligned} 
  x'(N) &= -3x - \frac{V'(\Phi)}{V(\Phi)}\sqrt{\frac{3}{2}} y^2 + \frac{3}{2}x \big[ 2x^2 + \gamma(1-x^2-y^2) \big] \\
  y'(N) &= \frac{V'(\Phi)}{V(\Phi)}\sqrt{\frac{3}{2}} xy + \frac{3}{2}y \big[ 2x^2 + \gamma(1-x^2-y^2) \big] \\
H'(N) &= -\frac{3}{2} H (2x^2 + \gamma(1-x^2-y^2)) \\
\Phi'(N) &= \sqrt{6} x}
{ \end{aligned} \right.
\end{equation}
where the time variable is $N = \log a$.

A simple case (which holds in LVS) is where the potential can be approximated by a single (steep) exponential,\footnote{This expression is valid far away from the minimum, where the uplifting term can be neglected. Including the uplift, the full potential takes the form
\begin{equation}
V(\Phi)= V_0 \big( (1-\varepsilon \Phi ^{3/2}) e^{- \lambda \Phi} + \delta e^{- \sqrt{6} \Phi }\big).
\end{equation}
The parameter $\delta$ needs to be fine-tuned to achieve a dS vacuum at $\Phi \sim \varepsilon^{-2/3}$. 
}
\be
V = V_0 \exp \left( - \lambda \frac{\Phi}{M_P} \right),
\ee 
so that $V'(\Phi)/ V(\Phi) = - \frac{\lambda}{M_P}$. 
For LVS, $\lambda = \sqrt{\frac{27}{2}}$. The precise value of $V_0$ will depend on the details of the compactification, but for reasonable values of $W_0$ we expect it to be of order $M_P^4$.
In this regime, the system \eqref{eq:xy2} is known to have a stable attractor solution where the scalar field and radiation have a fixed ratio of energy densities. The fixed point is characterised by
\begin{equation}
\Omega_k = x^2 = \frac{3}{2} \frac{\gamma^{2}}{\lambda^{2}}\quad \quad  \Omega_p = y^2 = \frac{3(2-\gamma) \gamma}{2 \lambda^{2}} \quad \quad  \Omega_{\gamma}= 1-x^2-y^2 = 1-\frac{3 \gamma}{\lambda^{2}}.
\end{equation}
If the attractor solution is obtained before the rolling field reaches the barrier, it will not overshoot.

\subsection{Picking up the Tracker Solution}

As a solution to the overshoot problem, the tracker
solution relies on other sources of radiation or matter `catching up' with the kinetic energy through their slower redshift. 
One aspect with this is that this takes (a lot of) time and so it may appear impractical for small seed amounts of radiation to grow sufficiently. 

A distinctive, and almost unique, feature of LVS (which makes it highly appealing for this purpose) is that
 the minimum of the potential is located a long way (in principle, many Planckian distances in field space) from the centre of moduli space where we assume inflation originally happened. While this feature has been touched on in \cite{Conlon:2008cj}, here we develop this aspect, and its phenomenological implications, significantly.
 
 As $a(t) \propto t^{1/3}$ and $\rho \propto a(t)^{-6}$ during 
 the kination epoch, the relative proportion of initial radiation $\rho_{\gamma}$ grows as
\be
\left( \frac{\rho_{\gamma}(t)}{\rho_{KE}(t)} \right)  = \left( \frac{t}{t_0} \right)^{2/3} \left( \frac{\rho_{\gamma}(t_0)}{\rho_{KE}(t_0)} \right),
\ee 
or equivalently,
\be
\ln \left( \frac{\rho_{\gamma}(t)}{\rho_{KE}(t)} \right) = \frac{2}{3} \ln \left( \frac{t}{t_0} \right) + \ln \left( \frac{\rho_{\gamma}(t_0)}{\rho_{KE}(t_0)} \right).
\ee
If kination-radiation equality is obtained, the system will find its way to the tracker solution as subsequent expansion sees radiation dominate over the kinetic energy; Hubble friction then effectively halts the field on its potential until the system evolves into the tracker solution. Kination-radiation equality occurs at
\be
\ln \left( \frac{t}{t_0} \right) = - \frac{3}{2} \ln \left( \frac{\rho_{\gamma}(t_0)}{\rho_{KE}(t_0)} \right).
\ee
As we would expect, the smaller the initial fraction of radiation, the longer it takes for this to be obtained. Using the field evolution Eq. (\ref{fieldkination}) during kination, the field displacement prior to reaching the tracker solution is
\be
\label{FieldChange}
\Delta \Phi = \langle \Phi \rangle - \langle \Phi_0 \rangle = \sqrt{\frac{3}{2}} \ln \left( \frac{\rho_{KE}(t_0)}{\rho_{\gamma}(t_0)} \right).
\ee
Expressed in terms of the physical compactification volume, this gives
\be
\label{VolumeChange}
\ln \left( \frac{{\mc V}_f}{{\mc V}_0} \right) = \frac{3}{2}  \ln \left( \frac{\rho_{KE}(t_0)}{\rho_{\gamma}(t_0)} \right).
\ee
The physical implications of Eqs. (\ref{FieldChange}) and (\ref{VolumeChange}) are striking. If we assume a small fraction of initial radiation 
$\ln \left( \frac{\rho_{\gamma}(t_0)}{\rho_{KE}(t_0)} \right)$, the field evolution is able to avoid overshooting and settle in the stable vacuum \emph{only provided the vacuum is sufficiently far away in field space}. The \emph{smaller} the initial fraction of radiation, the \emph{further} away the vacuum has to be in order for it to be accessible. As exponential potentials lead to a rapid decrease in vacuum energy along the roll, we also see that, somewhat paradoxically, vacua associated to much lower energy scales (and so with smaller barriers) become \emph{more} accessible not less.

This is one part of the inversion of conventional wisdom on overshoot within LVS. We now discuss the possible seeds for $\rho_{\gamma}(t_0)$ before returning to a discussion of physical and model-building implications.

\section{Seeds of radiation}\label{sc:rad}

Previous discussions of the use of radiation or matter to guide moduli towards the tracker solution have gone into relatively little detail on the origin and magnitude of the seed radiation. Here we discuss the possible sources of such seed radiation and their resulting phenomenological implications, with a particular focus on LVS.
While everything is model-dependent to some extent, there are generic reasons why radiation should be present.

\subsection{The Thermal de Sitter bath}\label{ssc:TdS}

The presumed dynamics is a primordial epoch of inflation exiting into a kination phase (although we wish to be agnostic about the actual model of inflation). If we suppose, consistent with the breakdown of slow-roll conditions at the end of inflation, that this transition is more or less instantaneous, any radiation present during inflation will still be present at the start of the kination phase.

While radiation is constantly diluted during inflation, this is balanced by the continual production of particles coming from the thermal de Sitter bath at temperature
\be
T_{dS} = \frac{H_{\rm{inf}}}{2 \pi}.
\ee
This provides a natural source of a radiation energy density
\be
\rho_{\gamma, dS} = \frac{ \pi^2}{30} g_{*} \left( \frac{H_{\rm{inf}}}{2 \pi} \right)^4,
\ee
where $H$ is the Hubble scale during inflation and $g^{*}$ is the effective number of massless degrees of freedom. There is no reason that $g^{*}$ should be small; both chiral fermions and gauge vector bosons naturally remain light during inflation. 
Even with scalars,
although the $\eta$ problem implies that it can be difficult to hold them lighter than the inflationary Hubble scale during inflation, this is only true if their mass is unprotected against quantum corrections. Axion fields -- which are generic in string theory -- are protected by their shift symmetry and remain naturally light during inflation. 
In a string theory context, $g_{*}$ could easily be rather large and $\mc{O}(100 - 1000)$, as the multiplicity of axions or gauge bosons may be set by the topology of the internal space and by Hodge numbers $h^{1,1}$ or $h^{2,1}$.

Setting $\rho_{KE}(t_0) = 3 H^2 M_P^2$, the thermal de Sitter bath would then lead to
\be
\label{thermalbath}
\frac{\rho_{\gamma}(t_0)}{\rho_{KE}(t_0)} = \frac{\pi^2}{90} g^{*} \left( \frac{H_{\rm{inf}}^2}{(2 \pi) M_P^2} \right).
\ee

\subsection{Perturbative Modulus `Decays'}\label{ssc:pd}

The thermal de Sitter bath has the advantage of universality: it treats all fields democratically. However, we expect there also to be radiation whose origin is specific to the fact that it is the volume modulus rolling down the exponential slope (analogous to the more conventional conversion of energy in the volume field to radiation at reheating). Although this may be small in \emph{overall} magnitude, it is competing with the similarly small thermal de Sitter bath (and the subsequent enhancement of radiation during the kination epoch makes even small initial quantities of radiation important).

We consider a perturbative origin for radiation from `decays' of a field rolling on a potential $V = V_0 \exp \left( - \lambda \frac{\Phi}{M_P} \right)$. Our approach is to model the potential as an (instantaneous) quadratic potential, and argue that for infinitesimal time periods it is reasonable to treat the field as equivalent to a conventional massive scalar particle within a quadratic potential, coupled to external degrees of freedom to which it can decay. 

Without loss of generality, we can consider the field starting at a value $\Phi = 0$ on the exponential potential $V = V_0 \exp \left( - \lambda \frac{\Phi}{M_P} \right)$.

At this time, we have
\bea
V & = & V_0, \\
\frac{dV}{d \Phi} & = & - \frac{\lambda}{M_P} V_0, \\
\frac{d^2 V}{d \Phi^2} & = & \left( \frac{\lambda}{M_P} \right)^2 V_0.
\eea
We wish to consider `instantaneous' dynamics, i.e. those over a time period $\Delta t \ll \frac{1}{H}$ for a field released at $\Phi = 0$ (from Eq. (\ref{HubbleEq2}), it follows that over such a period the field displacement is $\Delta \Phi \ll M_P$). To do so, we approximate the form of the potential in the immediate vicinity of $\Phi = 0$ as
\be
U(\Phi) = U_0 + \frac{1}{2} m^2 \left( \Phi - \Phi_0 \right)^2.
\ee
This is satisfied with $\Phi_0 = \frac{M_P}{\lambda}$, $U_0 = \frac{V_0}{2}$ and $m^2 = \left( \frac{\lambda}{M_P} \right)^2 V_0$. 

We now imagine, at the transition from the inflationary epoch, an approximately stationary field coming onto this potential at $\Phi = 0$. 
For a short period of time, we expect the field to `decay' to radiation in a similar fashion as if it were on the quadratic potential (instantaneously, it cannot `know the difference'). To model this source of initial radiation, we use the behaviour of a field starting at $\Phi = 0$ in the quadratic potential above.  How long does `instantaneous' last for? To give an estimate, we take approximately one Hubble time $\tau = \left( \frac{1}{H_0} \right) = \frac{\sqrt{3} M_P}{V_0^{1/2}}$.

We then compute the seed radiation content by regarding this field as a massive particle with
$$
m_{\Phi} = \left( \frac{\lambda}{M_P} \right) \sqrt{V_0},
$$
which decays via couplings to axions, photons or other light massless degrees of freedom. The precise couplings of LVS moduli to both matter and other dark particles, in the context of reheating, have been discussed in \cite{Cicoli:2012aq,Higaki:2012ar,Higaki:2012ba,Hebecker:2013lha,Higaki:2013lra, Angus:2014bia,Hebecker:2014gka,Cicoli:2015bpq,Hebecker:2022fcx,Baer:2022fou}. We expect a broadly similar analysis to hold here (although notice that these papers are working around the ultimate minimum of the scalar potential, while we are interested in couplings \emph{away} from the minimum).

While the numerical coefficients of the exact decay rate will be model-dependent, on general grounds we can write the decay rate $\Gamma_{\Phi \to X_i X_i}$ of any individual channel as
$$
\Gamma_i \equiv \Gamma_{\Phi \to X_i X_i} = \frac{\alpha_i}{16 \pi} \frac{m_{\Phi}^3}{M_P^2} = \frac{\alpha_i}{16 \pi} \left( \frac{\lambda}{M_P} \right)^3 \frac{V_0^{3/2}}{M_P^2}.
$$
In the absence of any additional suppression, $\alpha \sim 1$.
As the energy in the quadratic part of the potential (which we re-interpret as particles) is $V_0/2$ , it follows that the \emph{overall} fractional conversion rate of energy to radiation is $\Gamma_i/2$. As a naive estimate for the overall fraction of energy converted to radiation, we then have
\bea
\sum_i \frac{\Gamma}{2} \tau & = & \frac{\sqrt{3}}{2} \sum_i \alpha_i  \left( \frac{\lambda^3}{16 \pi} \right) \frac{V_0}{M_P^4} \\
& = & \frac{3 \sqrt{3}}{2} \sum_i \alpha_i  \left( \frac{\lambda^3}{16 \pi} \right) \frac{ H^2}{M_P^2},
\label{InitRad}
\eea
where the index $i$ runs over all decay channels. It is easily seen that this may be dominant over the thermal bath 
contribution Eq. (\ref{thermalbath}).

This formalism and in particular Eq. (\ref{InitRad}) also makes it clear that the dominant contribution to radiation comes from the immediate post-inflation period (due to the $\frac{H^2}{M_P^2}$ factor). As the field rolls down the exponential slope and picks up speed, the Hubble scale decreases and the rate at which seed radiation is generated falls off rapidly. 

\subsection{Gravitational Waves from Cosmic String Networks}

We now consider possibly the most interesting scenario for the generation of seed radiation.

This involves the radiation seed coming from gravitational waves emitted by cosmic string networks that were formed at the 
end of (brane) inflation. Such cosmic string networks can arise naturally in scenarios involving brane-antibrane inflation \cite{Dvali:1998pa,Burgess:2001fx,Dvali:2001fw,Sarangi:2002yt, Kachru:2003sx,Baumann:2007ah} (see \cite{Baumann:2014nda} for a comprehensive review). In these models, the end of inflation is coincident with brane-antibrane annihilation producing a large number of both fundamental and D-strings (D1-branes). In terms of their cosmic evolution, the most important quantity is the tension of such a string network, characterised by $G \mu$ (where $G$ is Newton's constant, $8 \pi G = \frac{1}{M_P^2}$). Reviews of cosmic strings can be found in \cite{Hindmarsh:1994re,Copeland:2009ga,Vachaspati:2015cma}.

Such string networks exhibit a scaling behaviour in which their energy density remains at a constant fraction of the overall energy density of the universe, 
\be
\rho_{string} \sim \mu H^2.
\ee 
This implies that as the universe expands, this energy in the string network decreases. This is not simply associated to a reduction in number density; the energy in the string network is lost via gravitational radiation emitted through string reconnections and from cusps. For this reason, cosmic string networks are regarded as one of the plausible sources of gravitational waves from the early universe (for example, see \cite{LIGOScientific:2017ikf} for a search and \cite{Caprini:2018mtu} for a review).

Such cosmic string networks can survive to today, where their tension is bounded via CMB constraints to be  $G \mu \lesssim 10^{-7}$. In the context of either fundamental or D-strings, the string tension $\mu \sim m_s^2$ and so observational consistency implies any fundamental string network would require either warping or a relatively low string scale.

We suppose a brane/antibrane inflationary model ending with the formation of a cosmic string network, which for simplicity we assume consists of vanilla fundamental strings (i.e. no warping). We also assume that the cosmic string scaling regime is attained rapidly and so soon after inflation the energy in the network is 
$\rho_{string} \sim \mu H_{\rm{inf}}^2$.
As the Hubble scale decreases the string network rapidly loses its energy via radiation from string cusps or reconnection. We assume an $\mc{O}(1)$ fraction of the string emission is to gravitational waves, implying an early presence of radiation
$$
\rho_{GW} \sim \mu H_{\rm{inf}}^2. 
$$
As $\mu \sim m_s^2$, we see that for a string scale $m_s \sim 10^{16} {\rm GeV} $ during inflation, giving $\mu \sim 10^{-4} M_P^2$, the resulting early fraction of injected radiation energy can be significant, e.g. 
$$
\frac{\rho_{GW}}{\rho_{total}} \sim 10^{-4}.
$$
This can easily dominate over either of the two previously identified sources of seed radiation Eqs. (\ref{thermalbath}) and (\ref{InitRad}).

Normally, such values of $G \mu$ would be immediately ruled out by the CMB constraint $G \mu \lesssim 10^{-7}$. However, here one appealing feature of LVS cosmology comes into play. During the long kination roll, the volume increases and
so -- as $m_s^2 \sim \frac{M_P^2}{\mc{V}}$ -- the cosmic string tension also decreases by a similar factor of $\frac{\mc{V}}{\mc{V}_0}$. The CMB constraints apply to the cosmic string tension today, not at the time of inflation -- and this increase in the volume reduces the string tension to a level entirely compatible with current observations.

In this scenario, the magnitude of the \emph{initial injection} of radiation as a proportion of the total energy density is set by $G \mu_{early}$,
while the \emph{current} energy density of the cosmic string network is set by $G \mu_{today}$ -- and these two can be very different.

This scenario also offers the possibility of a very intriguing aspect of the cosmological history of the universe. Suppose the initial injection of radiation from the cosmic string network was indeed dominantly in the form of gravitational waves. Then, as the universe approaches the tracker solution during the kination role, the radiation component consists of gravitational waves. While the universe is on the tracker solution, the radiation component makes up the largest individual part -- and so in fact, the dominant energy density of the universe is in the form of gravitational waves (for LVS parameters, $\Omega_{GW} = 19/27$)! 

This would hold during the tracker epoch. Once the modulus reaches the minimum and oscillates about it, the universe becomes matter-dominated and so this energy would then be redshifted away. Nonetheless, it would still be extremely striking if the universe did actually go through such a `graviation' epoch, where its energy was primarily in the form of gravitational waves. The possibility of such an epoch does not appear to have been considered very much in the literature, and it would be interesting to study its properties further.

\subsection{Primordial Black Holes}

Another interesting scenario for the initial seed radiation can occur if the dynamics at the end of inflation results in the production of primordial black holes (PBHs). Such black holes redshift as matter for as long as they remain black holes (i.e. before they evaporate). 
As $\rho_m \sim a(t)^{-3}$, if such PBHs were present the universe would evolves towards a matter tracker solution. 

As with cosmic strings, the production and initial density of black holes is highly model-dependent, as it crucially involves the spectrum of density perturbations in the universe and the details of the particular inflationary model (reviews of PBH can be found in e.g. \cite{Sasaki:2018dmp,Green:2020jor,Byrnes:2021jka}).
However, if PBHs are produced, their mass when produced is given by the horizon mass,
\be
M_{BH} \sim \frac{M_P^2}{H}.
\ee
From the point of production, the relative energy fraction in PBHs would increase throughout any kination epoch.
The subsequent evolution would depend on the mass of the primordial black hole. 
If the black holes all evaporated while still in the kination epoch, prior to the modulus reaching the LVS minimum, their mass-energy is converted to thermal radiation. In such a scenario, the role of the PBH has been solely to act as a radiation seed.

A more interesting scenario is when the PBHs live long enough that the universe reaches the `matter tracker'. 
At this point $\Omega_m = \frac{21}{27}$, and so the majority of the energy density of the universe would be in the form of primordial black holes. This will remain the case while the field settles into the minimum of the potential -- as the modulus itself also redshifts as matter, it cannot displace the black holes. Such an epoch, with energy density dominated by the mass of primordial black holes, would also be striking in itself as a feature of the (string) cosmological history of the universe.

If the PBHs were able to outlive the volume modulus, then reheating would end up proceeding via evaporation of the PBHs. Such an epoch would offer a potential solution to the problem of excessive dark radiation production in modulus decay. When the volume modulus decays, the branching ratio to axions appears to be $\mc{O}(1)$ as many of the Standard Model decay modes are suppressed \cite{Cicoli:2012aq,Higaki:2012ar}. However, for evaporation of a primordial black hole, modes should be more democratically accessible, potentially reducing the dark radiation fraction (depending on the number of hidden sectors).

However -- we do not expect this scenario to be viable, and would instead expect any PBHs to evaporate prior to the decay of the volume modulus. To see this, note that a black hole's lifetime behaves as $\tau_{BH} \propto \frac{M_{BH}^3}{M_P^4}$ whereas the lifetime of a modulus is $\tau_{\Phi} \sim \frac{M_P^2}{m_{\Phi^3}}$. We can therefore write the lifetime of a primordial black hole formed during kination as
\be
\tau_{BH} \sim \frac{M_P^2}{H_{form}^3},
\ee
where $H_{form}$ is the Hubble scale at the time of formation. However, as during the kination epoch $H > m_{\Phi}$ (as equality occurs when the field reaches the minimum of the potential), we find that 
\be
\tau_{BH} < \tau_{\Phi}
\ee
and so the black hole evaporates prior to modulus decay. It therefore appears unlikely that any black holes formed during the kination epoch could outlive the volume modulus.

\section{Model-Building Implications}\label{sec:ph}

Assuming that the kination phase starts directly at the end of inflation, one can use an assumption about the source of the seed radiation to relate the inflationary scale $\Lambda_{\rm{inf}}$ to the minimal field displacement required to settle into the minimum without overshooting.
 
From \eqref{VolumeChange}, kination-radiation equality is reached for 
\begin{equation}\label{eq:vmin}
\mathcal{V}_{\rm{eq}}=\Bigg( \frac{V_0}{\Lambda_{\rm{inf}}^4} \Bigg)^{\sqrt\frac{3}{2 \lambda^2}} \Omega_{\gamma,0}^{-3/2},
\end{equation}
where the initial volume has been rewritten in terms of $\Lambda_{\rm{inf}}$ and the parameters appearing in the potential. To avoid overshooting, we require the volume at the minimum to satisfy
$\mathcal{V}_{\rm{min}} \gtrsim \mathcal{V}_{\rm{eq}}$.

In the most minimal realisation (see section \ref{ssc:TdS}), the initial radiation will come from the 
thermal de Sitter radiation, viewed as a bath with $T_{\rm{dS}}= \frac{H}{2 \pi}$. All relativistic species with $m \ll \Lambda_{\rm{inf}}$ will contribute to a normalised energy density of
\begin{equation}
\Omega^0_{\rm{th}}= g^* \frac{\pi^2}{30} \Bigg( \frac{H_{\rm{inf}}}{2 \pi}\Bigg)^4  \frac{1}{3 H_{\rm{inf}}^2 M_P^2} = \frac{g^*}{270 \times 16 \pi^2} \Bigg( \frac{\Lambda_{\rm{inf}}}{M_P}\Bigg)^4.
\end{equation}
Under the conservative assumption that no new physics appears below the inflationary scale, the number of degrees of freedom would be $g^*=g^*_{\rm{SM}} \sim \mathcal{O}(10^2)$. Eq. \eqref{eq:vmin} implies
\begin{equation}\label{eq:blvs}
\mathcal{V}_{\rm{min}} \gtrsim 8.0 \times 10^{17} \,\,  \bigg( \frac{g^*}{100}\bigg)^{-3/2}
\Bigg(\frac{V_0}{M_P^4} \Bigg)^{1/3} \Bigg(\frac{\Lambda_{\rm{inf}}}{3 \cdot 10^{16}\, {\rm{GeV}}} \Bigg)^{-22/3},
\end{equation}
where we have used the LVS value of $\lambda = \sqrt{\frac{27}{2}}$. It would be entirely reasonable, however, to have $g^* =\mathcal{O}(10^3-10^4)$, potentially reducing the estimate \eqref{eq:blvs} by up to 3 orders of magnitude. As mentioned earlier, such a large number could easily arise from the high multiplicity of axions and gauge bosons in a typical compactification or from hidden sectors decoupled from the Standard Model.

Pushing this reasoning to its limits, one could imagine using the largest conceivable number of degrees of freedom in the UV, {\it{i.e.}} saturating the species bound of \cite{Dvali:2007hz,Dvali:2007wp}. If one assumes the inflationary scale to lie below the UV cutoff, this gives an absolute upper bound on the number of the EFT light degrees of freedom 
\begin{equation}\label{eq:thb}
g^* \lesssim \frac{M_P^2}{\Lambda_{\rm{inf}}^2}.
\end{equation}
Substituting in \eqref{eq:thb}, this turns into the laxer constraint
\begin{equation}\label{eq:llb}
\mathcal{V}_{\rm{min}} \gtrsim 1.6 \times 10^{15} \,\,
\Bigg(\frac{V_0}{M_P^4} \Bigg)^{1/3} \Bigg(\frac{\Lambda_{\rm{inf}}}{ 3 \cdot 10^{16}\, {\rm{GeV}} } \Bigg)^{-13/3}.
\end{equation}
While values of the volume appearing in \eqref{eq:llb} for realistic inflationary scales are certainly closer to what one would require for phenomenology, it is not clear in practice how and if such a high number of light states could arise. Therefore, we only interpret \eqref{eq:llb} as an absolute lower bound on the volume of vacua which can avoid overshooting through thermal dS radiation. In both cases, typical values for the volumes as a function of the inflationary scale are reported in Table \ref{tab:vol1}.

 %%%%%%%%%%%%%%%%%%%%%
\begin{table}[h!]
\centering
\begin{tabular}{|cl|c|ccc|}
\hline
&  & $\Lambda_{\rm{inf}} ( {\rm{GeV}} )$   & $ 3 \cdot 10^{16} $ & $3 \cdot 10^{15}$ & $ 3 \cdot 10^{14}$   \\
\hline
& $ g^*=10^2$ & $ \mathcal{V}_{\rm{min}}$  & $ 8.0 \cdot 10^{17} $ & $ 1.7 \cdot 10^{25} $ & $ 3.7 \cdot 10^{32} $   \\
& $g^*=10^4$ & $ \mathcal{V}_{\rm{min}}$  & $ 8.0 \cdot 10^{14} $ & $ 1.7 \cdot 10^{22} $ & $ 3.7 \cdot 10^{29} $   \\
\hdashline
& $g^*=M_P^2/\Lambda_{\rm{inf}}^2$ & $\mathcal{V}_{\rm{min}}$  & $ 1.6 \cdot 10^{15}$ & $ 3.3 \cdot 10^{19} $ & $ 7.2 \cdot 10^{23} $   \\
\hline
\end{tabular}
\caption{\it {Lower bound for a stable volume (measured in units of $(2 \pi \sqrt{\alpha'})^6$}) in LVS as a function of the inflation scale, assuming only a thermal de Sitter radiation bath. The first line conservatively assumes the number of relativistic degrees of freedom after inflation to be $g^* = g_{\rm{SM}} \simeq 100$, while the second one uses the more aggressive estimate $g^* = 10^4$. The third line gives an absolute lower bound assuming the maximum number of degrees of freedom compatible with the species conjecture.}
\label{tab:vol1}
\end{table}
%%%%%%%%%%%%%%%%%%%%%

In the case of perturbative decays, treated in section \ref{ssc:pd}, the parametric scaling of the initial radiation density as a function of the Hubble scale is the same as for thermal de Sitter radiation ($\propto H^4$). However, the numerical coefficient in front of \eqref{InitRad} gives a larger enhancement:
\begin{equation}
\Omega^0_{\rm{dec}}= \frac{3 \sqrt{3}}{2} \sum_i \alpha_i  \left( \frac{\lambda^3}{16 \pi} \right) \frac{ H^2}{M_P^2} = \frac{9 \sqrt{3}}{2} g_{\rm{dec}}  \left( \frac{\lambda^3}{16 \pi} \right) \Bigg( \frac{\Lambda_{\rm{inf}}}{M_P}\Bigg)^4.
\end{equation}
The factor $g_{\rm{dec}}$ is the effective number of species to which the volume modulus can decay, taking a value $\alpha_i=1$. Then, taking again $\lambda = \sqrt{\frac{27}{2}}$, there is a relative factor
\begin{equation}
\frac{\Omega^0_{\rm{dec}}}{\Omega^0_{\rm{th}}}  \simeq  3.3 \cdot 10^{5} \frac{g_{\rm{dec}}}{g^*}.
\end{equation}
Then,
\begin{equation}\label{eq:blvst}
\mathcal{V} \gtrsim 4.2 \times 10^{9} \,\,
\,\,  \bigg( \frac{g_{\rm{dec}}}{100}\bigg)^{-3/2}
\Bigg(\frac{V_0}{M_P^4} \Bigg)^{1/3} \Bigg(\frac{\Lambda_{\rm{inf}}}{ 3 \cdot 10^{16}\, {\rm{GeV}}} \Bigg)^{-22/3}.
\end{equation}
Typical values are summarised in Table \ref{tab:vol2}.

 %%%%%%%%%%%%%%%%%%%%%
\begin{table}[h!]
\centering
\begin{tabular}{|cl|c|cccc|}
\hline
&  & $\Lambda_{\rm{inf}} ( {\rm{GeV}} )$   & $ 3 \cdot 10^{16} $ & $3 \cdot 10^{15}$ & $ 3 \cdot 10^{14}$  & $ 3 \cdot 10^{13}$  \\
\hline
& $ g_{\rm{dec}}=1$ & $ \mathcal{V}_{\rm{min}}$  & $ 4.1 \cdot 10^{12} $ & $ 8.9 \cdot 10^{19} $ & $ 1.9 \cdot 10^{27} $ & $ 4.1 \cdot 10^{34} $ \\
& $ g_{\rm{dec}}=10^2$ & $ \mathcal{V}_{\rm{min}}$  & $ 4.1 \cdot 10^{9} $ & $ 8.9 \cdot 10^{16} $ & $ 1.9 \cdot 10^{24} $ & $ 4.1 \cdot 10^{31} $ \\
& $ g_{\rm{dec}}=10^4$ & $ \mathcal{V}_{\rm{min}}$  & $ 4.1 \cdot 10^{6} $ & $ 8.9 \cdot 10^{13} $ & $ 1.9 \cdot 10^{21} $ & $ 4.1 \cdot 10^{28} $ \\
\hline
\end{tabular}
\caption{\it {Lower bound for the volume (measured in units of $(2 \pi \sqrt{\alpha'})^6$})  in LVS as a function of the inflation scale, assuming initial radiation from perturbative decays. The first line conservatively assumes the number of light degrees of freedom the volume can decay to after inflation to be $g_{\rm{dec}} = 1$, while the second and third assume the more aggressive estimates $g_{\rm{dec}} = g_{\rm{SM}} \simeq 100$ and $g_{\rm{dec}} = 10^4$. }
\label{tab:vol2}
\end{table}
%%%%%%%%%%%%%%%%%%%%%

Perhaps the most interesting scenario of seed radiation involved the production of gravitational waves from cosmic string scaling networks.
Our results therefore motivate the detailed study of inflationary models within LVS which end in brane/antibrane annihilation and the production of cosmic strings. So far, the most detailed analyses of brane inflation models have been carried out within KKLT constructions. LVS offers a very different phenomenology, in particular the prospect of a large reduction in the value of the string scale between the end of inflation and the present day. The low string scales in LVS would also make a fundamental cosmological superstring network observationally viable, with the tension naturally satisfying $G \mu \lesssim 10^{-7}$. 

In terms of avoiding overshoot, there are two important general points to make. The first is the fact that in all the possible radiation seeds we have considered, the larger the initial ratio of $\frac{H_{\rm{inf}}}{M_P}$, the larger the initial radiation fraction, and so the easier it is subsequently to obtain kination-radiation equality and avoid an overshoot by locating, and switching onto, the tracker solution. This is also true for cosmic strings and PBHs, although estimates such as \eqref{eq:blvs} or \eqref {eq:blvst} would be more model dependent in those cases. 

The second point is more LVS-specific: this is that the location of the minimum at a large distance in field space away from the `center' of moduli space is crucial for the presence of hierarchical scale separation and, when using these hierarchies for particle physics, crucial for explaining the large ratio between the Planck and electroweak scales. In this context, the vacuum can only be located (avoiding overshooting) if the inflationary scale is sufficiently high.

This inverts the usual logic of the overshoot problem -- in this framing, the problem becomes \emph{less} severe the higher up on the potential we start. For those who find anthropic arguments appealing, in this context we can also turn this into an anthropic argument for the existence of a large hierarchy between the scale of inflation (i.e. where we start on the roll) and the electroweak scale (characterised by the location of the minimum at a large distance in field space from the inflationary locus). Indeed, the gravitino mass - which could be thought to provide the scale of SUSY breaking - scales as
\begin{equation}
m_{3/2} \simeq\frac{M_P |W_0|}{\mathcal{V} }
\end{equation}
in LVS. The relationship of the gravitino mass and MSSM soft terms is subtle in LVS; while original analyses \cite{cqs} assumed these would be comparable, more careful analyses of soft terms and sequestering \cite{Blumenhagen:2009gk,Reece:2015qbf} show that we would actually expect soft terms at a scale
\be
M_{\rm{soft}} \sim \frac{m_{3/2}^{3/2}}{M_P^{1/2}}.
\ee
In any case, under the hypothesis that the solution to the hierarchy problem is connected in some way to the presence of low energy supersymmetry, this gives a direct link between large volumes and a low electroweak scale.
Depending on the precise model details, the appropriate LVS volume to match onto particle physics may be $10^{6} \lesssim \mc{V} \lesssim 10^{13}$. From the estimates in Tables \ref{tab:vol1} and \ref{tab:vol2}, these volumes may reasonably be large enough to avoid overshooting but with a clear requirement for relatively high-scale inflation (not necessarily transPlanckian field excursions).

\section{Conclusions}
\label{sec:Conclusions}

String theorists should feel their ears burning and exhibit a professional interest whenever fields undergo trans-Planckian displacements $\Delta \Phi \gtrsim M_P$. This fact is widely acknowledged for models of large-field inflation. Less attention, however, has been paid to the case of a possible kination epoch in the early universe.

The overshoot problem, and tracker solutions as a means to solve it, have over the years received intermittent attention in string cosmology.
In the context of LVS, we have shown here that a long kination epoch is simultaneously a natural feature of cosmology in LVS and an attractive method of solving the overshoot problem: initial seed radiation from the immediate end of the inflationary epoch grows in relative importance during kination, allowing the rolling field to locate the tracker solution and avoid overshooting the minimum. 

This approach requires the initial seed radiation to grow from a (presumed) small value to 
reach an order unity fraction of the overall energy density. 
There are both universal and model-dependent options for the origin of the original seed radiation, but one aspect they  have in common is that the initial radiation fraction grows with positive powers of $\left( \frac{H_{\rm{inf}}}{M_P} \right)$. As it takes considerable time for the small initial radiation fraction to `catch up', 
the avoidance of overshoot is therefore made \emph{easier} by large inflationary scales and/or a large (significantly transPlanckian) field displacement between the inflationary epoch and the final vacuum.
This takes some time; LVS is rather distinctive here 
as it naturally contains a large trans-Planckian displacement in the field space between the locus of inflation and the final minimum. In the context of the different scales in LVS, this offers an anthropic argument for a large hierarchy between the weak and inflationary scales.

One particularly interesting scenario is when inflation ends with brane/anti-brane annihilation and the creation of a cosmic superstring network. As the string scale decreases during the kination roll, the network can be formed with a relatively high tension (far above bounds based on the universe today). If the network dissipates energy primarily through gravitational waves, this can lead to the universe passing through an epoch where its energy density is principally in the form of gravitational waves. This can also lead to a surviving network of fundamental cosmic superstrings; as phenomenological scenarios of LVS are expected to operate with low string scales, $10^{12} {\rm GeV} \lesssim m_s \lesssim 10^{16} {\rm GeV}$, the string tension $G \mu$ can naturally lie in interesting observational windows $10^{-7} \lesssim G\mu \lesssim 10^{-11}$.

It will be interesting to study the physics and cosmology of these scenarios further.

\section*{Acknowledgements}

We would like to thank Prateek Agrawal, Ed Hardy, Francesco Muia and Fernando Quevedo for discussions, and Bruno Bento for comments on the manuscript. FR is supported by the Dalitz Graduate Scholarship, jointly established by the Oxford University Department of Physics and Wadham College. We also thank STFC for funding via Consolidated Grant ST/T000864/1.

%\bibliography{bibabstract}

\begin{thebibliography}{10}

\bibitem{BaumannBook}
D.~Baumann, \emph{Cosmology}. CUP, 2022.

\bibitem{kklt}
S.~Kachru, R.~Kallosh, A.~D. Linde and S.~P. Trivedi, \emph{{De Sitter vacua in
  string theory}},
  \href{https://doi.org/10.1103/PhysRevD.68.046005}{\emph{Phys. Rev. D}
  {\bfseries 68} (2003) 046005}
  [\href{https://arxiv.org/abs/hep-th/0301240}{{\ttfamily hep-th/0301240}}].

\bibitem{bbcq}
V.~Balasubramanian, P.~Berglund, J.~P. Conlon and F.~Quevedo,
  \emph{{Systematics of moduli stabilisation in Calabi-Yau flux
  compactifications}},
  \href{https://doi.org/10.1088/1126-6708/2005/03/007}{\emph{JHEP} {\bfseries
  03} (2005) 007} [\href{https://arxiv.org/abs/hep-th/0502058}{{\ttfamily
  hep-th/0502058}}].

\bibitem{cqs}
J.~P. Conlon, F.~Quevedo and K.~Suruliz, \emph{{Large-volume flux
  compactifications: Moduli spectrum and D3/D7 soft supersymmetry breaking}},
  \href{https://doi.org/10.1088/1126-6708/2005/08/007}{\emph{JHEP} {\bfseries
  08} (2005) 007} [\href{https://arxiv.org/abs/hep-th/0505076}{{\ttfamily
  hep-th/0505076}}].

\bibitem{Kallosh:2004yh}
R.~Kallosh and A.~D. Linde, \emph{{Landscape, the scale of SUSY breaking, and
  inflation}}, \href{https://doi.org/10.1088/1126-6708/2004/12/004}{\emph{JHEP}
  {\bfseries 12} (2004) 004}
  [\href{https://arxiv.org/abs/hep-th/0411011}{{\ttfamily hep-th/0411011}}].

\bibitem{Blanco-Pillado:2005arr}
J.~J. Blanco-Pillado, R.~Kallosh and A.~D. Linde, \emph{{Supersymmetry and
  stability of flux vacua}},
  \href{https://doi.org/10.1088/1126-6708/2006/05/053}{\emph{JHEP} {\bfseries
  05} (2006) 053} [\href{https://arxiv.org/abs/hep-th/0511042}{{\ttfamily
  hep-th/0511042}}].

\bibitem{Kallosh:2007wm}
R.~Kallosh and A.~D. Linde, \emph{{Testing String Theory with CMB}},
  \href{https://doi.org/10.1088/1475-7516/2007/04/017}{\emph{JCAP} {\bfseries
  04} (2007) 017} [\href{https://arxiv.org/abs/0704.0647}{{\ttfamily
  0704.0647}}].

\bibitem{Conlon:2008cj}
J.~P. Conlon, R.~Kallosh, A.~D. Linde and F.~Quevedo, \emph{{Volume Modulus
  Inflation and the Gravitino Mass Problem}},
  \href{https://doi.org/10.1088/1475-7516/2008/09/011}{\emph{JCAP} {\bfseries
  09} (2008) 011} [\href{https://arxiv.org/abs/0806.0809}{{\ttfamily
  0806.0809}}].

\bibitem{BrusteinSteinhardt}
R.~Brustein and P.~J. Steinhardt, \emph{{Challenges for superstring
  cosmology}}, \href{https://doi.org/10.1016/0370-2693(93)90384-T}{\emph{Phys.
  Lett. B} {\bfseries 302} (1993) 196}
  [\href{https://arxiv.org/abs/hep-th/9212049}{{\ttfamily hep-th/9212049}}].

\bibitem{German:2001tz}
G.~German, G.~G. Ross and S.~Sarkar, \emph{{Low scale inflation}},
  \href{https://doi.org/10.1016/S0550-3213(01)00258-9}{\emph{Nucl. Phys. B}
  {\bfseries 608} (2001) 423}
  [\href{https://arxiv.org/abs/hep-ph/0103243}{{\ttfamily hep-ph/0103243}}].

\bibitem{hep-th/0605264}
H.~Ooguri and C.~Vafa, \emph{{On the Geometry of the String Landscape and the
  Swampland}},
  \href{https://doi.org/10.1016/j.nuclphysb.2006.10.033}{\emph{Nucl. Phys. B}
  {\bfseries 766} (2007) 21}
  [\href{https://arxiv.org/abs/hep-th/0605264}{{\ttfamily hep-th/0605264}}].

\bibitem{181005506}
H.~Ooguri, E.~Palti, G.~Shiu and C.~Vafa, \emph{{Distance and de Sitter
  Conjectures on the Swampland}},
  \href{https://doi.org/10.1016/j.physletb.2018.11.018}{\emph{Phys. Lett. B}
  {\bfseries 788} (2019) 180}
  [\href{https://arxiv.org/abs/1810.05506}{{\ttfamily 1810.05506}}].
  
  \bibitem{Silverstein:2008sg}
E.~Silverstein and A.~Westphal,
\emph{{Monodromy in the CMB: Gravity Waves and String Inflation,}}
\href{https://doi.org/10.1103/PhysRevD.78.106003}{\emph{Phys. Rev. D}
  {\bfseries 78} (2008) 106003}
  [\href{https://arxiv.org/abs/0803.3085}{{\ttfamily 0803.3085}}].
  
\bibitem{McAllister:2008hb}
L.~McAllister, E.~Silverstein and A.~Westphal,
\emph{{Gravity Waves and Linear Inflation from Axion Monodromy,}}
\href{https://doi.org/10.1103/PhysRevD.82.046003}{\emph{Phys. Rev. D}
  {\bfseries 82} (2010) 046003}
  [\href{https://arxiv.org/abs/0808.0706}{{\ttfamily 0808.0706}}].


\bibitem{Wetterich:1987fm}
C.~Wetterich, \emph{{Cosmology and the Fate of Dilatation Symmetry}},
  \href{https://doi.org/10.1016/0550-3213(88)90193-9}{\emph{Nucl. Phys. B}
  {\bfseries 302} (1988) 668}
  [\href{https://arxiv.org/abs/1711.03844}{{\ttfamily 1711.03844}}].

\bibitem{Copeland:1997et}
E.~J. Copeland, A.~R. Liddle and D.~Wands, \emph{{Exponential potentials and
  cosmological scaling solutions}},
  \href{https://doi.org/10.1103/PhysRevD.57.4686}{\emph{Phys. Rev. D}
  {\bfseries 57} (1998) 4686}
  [\href{https://arxiv.org/abs/gr-qc/9711068}{{\ttfamily gr-qc/9711068}}].

\bibitem{Ferreira:1997hj}
P.~G. Ferreira and M.~Joyce, \emph{{Cosmology with a primordial scaling
  field}}, \href{https://doi.org/10.1103/PhysRevD.58.023503}{\emph{Phys. Rev.
  D} {\bfseries 58} (1998) 023503}
  [\href{https://arxiv.org/abs/astro-ph/9711102}{{\ttfamily
  astro-ph/9711102}}].

\bibitem{Barreiro:1998aj}
T.~Barreiro, B.~de~Carlos and E.~J. Copeland, \emph{{Stabilizing the dilaton in
  superstring cosmology}},
  \href{https://doi.org/10.1103/PhysRevD.58.083513}{\emph{Phys. Rev. D}
  {\bfseries 58} (1998) 083513}
  [\href{https://arxiv.org/abs/hep-th/9805005}{{\ttfamily hep-th/9805005}}].
  
  \bibitem{hepth0001112}
G.~Huey, P.~J.~Steinhardt, B.~A.~Ovrut and D.~Waldram, \emph{{A Cosmological mechanism for stabilizing moduli}},
  \href{https://doi.org/10.1016/S0370-2693(00)00152-0}{\emph{Phys. Lett. B}
  {\bfseries476} (2000), 379-386}
   [\href{https://arxiv.org/abs/hep-th/0001112}{{\ttfamily hep-th/0001112}}].
  
 
  \bibitem{hepph0010102}
T.~Barreiro, B.~de Carlos and N.~J.~Nunes, \emph{{Moduli evolution in heterotic scenarios}},
  \href{https://doi.org/10.1016/S0370-2693(00)01308-3}{\emph{Phys. Lett. B}
  {\bfseries 497} (2001), 136-144}
 [\href{https://arxiv.org/abs/hep-ph/0010102}{{\ttfamily hep-ph/0010102}}].
  

\bibitem{hepth0408160}
R.~Brustein, S.~P.~de Alwis and P.~Martens,\emph{{Cosmological stabilization of moduli with steep potentials}},
  \href{https://journals.aps.org/prd/abstract/10.1103/PhysRevD.70.126012}
{\emph{Phys. Rev.
  D} {\bfseries 70} (2004), 126012}
  [\href{https://arxiv.org/abs/hep-th/0408160}{{\ttfamily hep-th/0408160}}].



\bibitem{hepph0506045}
T.~Barreiro, B.~de Carlos, E.~Copeland and N.~J.~Nunes,\emph{{Moduli evolution in the presence of flux compactifications}},
  \href{https://journals.aps.org/prd/abstract/10.1103/PhysRevD.72.106004}
{\emph{Phys. Rev.
  D} {\bfseries 72} (2005), 106004}
  [\href{https://arxiv.org/abs/hep-ph/0506045}{{\ttfamily hep-ph/0506045}}].


\bibitem{Cicoli:2012aq}
M.~Cicoli, J.~P. Conlon and F.~Quevedo, \emph{{Dark radiation in LARGE volume
  models}}, \href{https://doi.org/10.1103/PhysRevD.87.043520}{\emph{Phys. Rev.
  D} {\bfseries 87} (2013) 043520}
  [\href{https://arxiv.org/abs/1208.3562}{{\ttfamily 1208.3562}}].

\bibitem{Higaki:2012ar}
T.~Higaki and F.~Takahashi, \emph{{Dark Radiation and Dark Matter in Large
  Volume Compactifications}},
  \href{https://doi.org/10.1007/JHEP11(2012)125}{\emph{JHEP} {\bfseries 11}
  (2012) 125} [\href{https://arxiv.org/abs/1208.3563}{{\ttfamily 1208.3563}}].

\bibitem{Higaki:2012ba}
T.~Higaki, K.~Kamada and F.~Takahashi, \emph{{Higgs, Moduli Problem,
  Baryogenesis and Large Volume Compactifications}},
  \href{https://doi.org/10.1007/JHEP09(2012)043}{\emph{JHEP} {\bfseries 09}
  (2012) 043} [\href{https://arxiv.org/abs/1207.2771}{{\ttfamily 1207.2771}}].

\bibitem{Hebecker:2013lha}
A.~Hebecker, A.~K. Knochel and T.~Weigand, \emph{{The Higgs mass from a
  String-Theoretic Perspective}},
  \href{https://doi.org/10.1016/j.nuclphysb.2013.05.004}{\emph{Nucl. Phys. B}
  {\bfseries 874} (2013) 1} [\href{https://arxiv.org/abs/1304.2767}{{\ttfamily
  1304.2767}}].

\bibitem{Higaki:2013lra}
T.~Higaki, K.~Nakayama and F.~Takahashi, \emph{{Moduli-Induced Axion Problem}},
  \href{https://doi.org/10.1007/JHEP07(2013)005}{\emph{JHEP} {\bfseries 07}
  (2013) 005} [\href{https://arxiv.org/abs/1304.7987}{{\ttfamily 1304.7987}}].

\bibitem{Angus:2014bia}
S.~Angus, \emph{{Dark Radiation in Anisotropic LARGE Volume
  Compactifications}},
  \href{https://doi.org/10.1007/JHEP10(2014)184}{\emph{JHEP} {\bfseries 10}
  (2014) 184} [\href{https://arxiv.org/abs/1403.6473}{{\ttfamily 1403.6473}}].

\bibitem{Hebecker:2014gka}
A.~Hebecker, P.~Mangat, F.~Rompineve and L.~T. Witkowski, \emph{{Dark Radiation
  predictions from general Large Volume Scenarios}},
  \href{https://doi.org/10.1007/JHEP09(2014)140}{\emph{JHEP} {\bfseries 09}
  (2014) 140} [\href{https://arxiv.org/abs/1403.6810}{{\ttfamily 1403.6810}}].

\bibitem{Cicoli:2015bpq}
M.~Cicoli and F.~Muia, \emph{{General Analysis of Dark Radiation in Sequestered
  String Models}}, \href{https://doi.org/10.1007/JHEP12(2015)152}{\emph{JHEP}
  {\bfseries 12} (2015) 152}
  [\href{https://arxiv.org/abs/1511.05447}{{\ttfamily 1511.05447}}].

\bibitem{Hebecker:2022fcx}
A.~Hebecker, J.~Jaeckel and M.~Wittner, \emph{{Axions in String Theory and the
  Hydra of Dark Radiation}},
  \href{https://arxiv.org/abs/2203.08833}{{\ttfamily 2203.08833}}.

\bibitem{Baer:2022fou}
H.~Baer, V.~Barger and R.~W. Deal, \emph{{On dark radiation from string moduli
  decay to ALPs}},
  \href{https://doi.org/10.1016/j.jheap.2022.04.001}{\emph{JHEAp} {\bfseries
  34} (2022) 40} [\href{https://arxiv.org/abs/2204.01130}{{\ttfamily
  2204.01130}}].

\bibitem{Dvali:1998pa}
G.~R. Dvali and S.~H.~H. Tye, \emph{{Brane inflation}},
  \href{https://doi.org/10.1016/S0370-2693(99)00132-X}{\emph{Phys. Lett. B}
  {\bfseries 450} (1999) 72}
  [\href{https://arxiv.org/abs/hep-ph/9812483}{{\ttfamily hep-ph/9812483}}].

\bibitem{Burgess:2001fx}
C.~P. Burgess, M.~Majumdar, D.~Nolte, F.~Quevedo, G.~Rajesh and R.-J. Zhang,
  \emph{{The Inflationary brane anti-brane universe}},
  \href{https://doi.org/10.1088/1126-6708/2001/07/047}{\emph{JHEP} {\bfseries
  07} (2001) 047} [\href{https://arxiv.org/abs/hep-th/0105204}{{\ttfamily
  hep-th/0105204}}].

\bibitem{Dvali:2001fw}
G.~R. Dvali, Q.~Shafi and S.~Solganik, \emph{{D-brane inflation}},  in
  \emph{{4th European Meeting From the Planck Scale to the Electroweak Scale}},
  5, 2001, \href{https://arxiv.org/abs/hep-th/0105203}{{\ttfamily
  hep-th/0105203}}.

\bibitem{Sarangi:2002yt}
S.~Sarangi and S.~H.~H. Tye, \emph{{Cosmic string production towards the end of
  brane inflation}},
  \href{https://doi.org/10.1016/S0370-2693(02)01824-5}{\emph{Phys. Lett. B}
  {\bfseries 536} (2002) 185}
  [\href{https://arxiv.org/abs/hep-th/0204074}{{\ttfamily hep-th/0204074}}].

\bibitem{Kachru:2003sx}
S.~Kachru, R.~Kallosh, A.~D. Linde, J.~M. Maldacena, L.~P. McAllister and S.~P.
  Trivedi, \emph{{Towards inflation in string theory}},
  \href{https://doi.org/10.1088/1475-7516/2003/10/013}{\emph{JCAP} {\bfseries
  10} (2003) 013} [\href{https://arxiv.org/abs/hep-th/0308055}{{\ttfamily
  hep-th/0308055}}].

\bibitem{Baumann:2007ah}
D.~Baumann, A.~Dymarsky, I.~R. Klebanov and L.~McAllister, \emph{{Towards an
  Explicit Model of D-brane Inflation}},
  \href{https://doi.org/10.1088/1475-7516/2008/01/024}{\emph{JCAP} {\bfseries
  01} (2008) 024} [\href{https://arxiv.org/abs/0706.0360}{{\ttfamily
  0706.0360}}].

\bibitem{Baumann:2014nda}
D.~Baumann and L.~McAllister, \emph{{Inflation and String Theory}}, Cambridge
  Monographs on Mathematical Physics. Cambridge University Press, 5, 2015,
  \href{https://doi.org/10.1017/CBO9781316105733}{10.1017/CBO9781316105733},
  [\href{https://arxiv.org/abs/1404.2601}{{\ttfamily 1404.2601}}].

\bibitem{Hindmarsh:1994re}
M.~B. Hindmarsh and T.~W.~B. Kibble, \emph{{Cosmic strings}},
  \href{https://doi.org/10.1088/0034-4885/58/5/001}{\emph{Rept. Prog. Phys.}
  {\bfseries 58} (1995) 477}
  [\href{https://arxiv.org/abs/hep-ph/9411342}{{\ttfamily hep-ph/9411342}}].

\bibitem{Copeland:2009ga}
E.~J. Copeland and T.~W.~B. Kibble, \emph{{Cosmic Strings and Superstrings}},
  \href{https://doi.org/10.1098/rspa.2009.0591}{\emph{Proc. Roy. Soc. Lond. A}
  {\bfseries 466} (2010) 623}
  [\href{https://arxiv.org/abs/0911.1345}{{\ttfamily 0911.1345}}].

\bibitem{Vachaspati:2015cma}
T.~Vachaspati, L.~Pogosian and D.~Steer, \emph{{Cosmic Strings}},
  \href{https://doi.org/10.4249/scholarpedia.31682}{\emph{Scholarpedia}
  {\bfseries 10} (2015) 31682}
  [\href{https://arxiv.org/abs/1506.04039}{{\ttfamily 1506.04039}}].

\bibitem{LIGOScientific:2017ikf}
{\scshape LIGO Scientific, Virgo} collaboration, B.~P. Abbott et~al.,
  \emph{{Constraints on cosmic strings using data from the first Advanced LIGO
  observing run}},
  \href{https://doi.org/10.1103/PhysRevD.97.102002}{\emph{Phys. Rev. D}
  {\bfseries 97} (2018) 102002}
  [\href{https://arxiv.org/abs/1712.01168}{{\ttfamily 1712.01168}}].

\bibitem{Caprini:2018mtu}
C.~Caprini and D.~G. Figueroa, \emph{{Cosmological Backgrounds of Gravitational
  Waves}}, \href{https://doi.org/10.1088/1361-6382/aac608}{\emph{Class. Quant.
  Grav.} {\bfseries 35} (2018) 163001}
  [\href{https://arxiv.org/abs/1801.04268}{{\ttfamily 1801.04268}}].

\bibitem{Sasaki:2018dmp}
M.~Sasaki, T.~Suyama, T.~Tanaka and S.~Yokoyama, \emph{{Primordial black
  holes\textemdash{}perspectives in gravitational wave astronomy}},
  \href{https://doi.org/10.1088/1361-6382/aaa7b4}{\emph{Class. Quant. Grav.}
  {\bfseries 35} (2018) 063001}
  [\href{https://arxiv.org/abs/1801.05235}{{\ttfamily 1801.05235}}].

\bibitem{Green:2020jor}
A.~M. Green and B.~J. Kavanagh, \emph{{Primordial Black Holes as a dark matter
  candidate}}, \href{https://doi.org/10.1088/1361-6471/abc534}{\emph{J. Phys.
  G} {\bfseries 48} (2021) 043001}
  [\href{https://arxiv.org/abs/2007.10722}{{\ttfamily 2007.10722}}].

\bibitem{Byrnes:2021jka}
C.~T. Byrnes and P.~S. Cole, \emph{{Lecture notes on inflation and primordial
  black holes}},  12, 2021, \href{https://arxiv.org/abs/2112.05716}{{\ttfamily
  2112.05716}}.

\bibitem{Dvali:2007hz}
G.~Dvali, \emph{{Black Holes and Large N Species Solution to the Hierarchy
  Problem}}, \href{https://doi.org/10.1002/prop.201000009}{\emph{Fortsch.
  Phys.} {\bfseries 58} (2010) 528}
  [\href{https://arxiv.org/abs/0706.2050}{{\ttfamily 0706.2050}}].

\bibitem{Dvali:2007wp}
G.~Dvali and M.~Redi, \emph{{Black Hole Bound on the Number of Species and
  Quantum Gravity at LHC}},
  \href{https://doi.org/10.1103/PhysRevD.77.045027}{\emph{Phys. Rev. D}
  {\bfseries 77} (2008) 045027}
  [\href{https://arxiv.org/abs/0710.4344}{{\ttfamily 0710.4344}}].

\bibitem{Blumenhagen:2009gk}
R.~Blumenhagen, J.~P. Conlon, S.~Krippendorf, S.~Moster and F.~Quevedo,
  \emph{{SUSY Breaking in Local String/F-Theory Models}},
  \href{https://doi.org/10.1088/1126-6708/2009/09/007}{\emph{JHEP} {\bfseries
  09} (2009) 007} [\href{https://arxiv.org/abs/0906.3297}{{\ttfamily
  0906.3297}}].

\bibitem{Reece:2015qbf}
M.~Reece and W.~Xue, \emph{{SUSY\textquoteright{}s Ladder: reframing
  sequestering at Large Volume}},
  \href{https://doi.org/10.1007/JHEP04(2016)045}{\emph{JHEP} {\bfseries 04}
  (2016) 045} [\href{https://arxiv.org/abs/1512.04941}{{\ttfamily
  1512.04941}}].

\end{thebibliography}
\bibliographystyle{JHEP}

\providecommand{\href}[2]{#2}\begingroup\raggedright\endgroup

\end{document}